\documentclass[twocolumn]{aastex6}
\usepackage{graphicx}
\usepackage{float}
\usepackage{amsmath}
\usepackage{amssymb}
\usepackage{textcomp}
\usepackage[caption=false]{subfig}
\begin{document}

\title{Mapping the Tidal Destruction of the Hercules Dwarf: A Wide-Field DECam Imaging Search for RR Lyrae Stars}

\author{Christopher Garling\altaffilmark{1,2}}
\altaffiltext{1}{Department of Astronomy, The Ohio State University, 140 W. 18th Avenue, Columbus, OH 43210, USA; garling.14@osu.edu}
\altaffiltext{2}{Department of Astronomy, Haverford College, 370 Lancaster Avenue, Haverford, PA 19041, USA}

\author{Beth Willman\altaffilmark{3,4}}
\altaffiltext{3}{Steward Observatory, University of Arizona, 933 N. Cherry Avenue, Tucson, AZ 85721, USA}
\altaffiltext{4}{LSST, University of Arizona, 933 N. Cherry Avenue, Tucson, AZ 85721, USA}

\author{David J. Sand\altaffilmark{3}}

\author{Jonathan Hargis\altaffilmark{5}}
\altaffiltext{5}{Space Telescope Science Institute, 3700 San Martin Drive, Baltimore, MD 21218, USA}

\author{Denija Crnojevi{\'{c}}\altaffilmark{6}}
\altaffiltext{6}{Department of Physics and Astronomy, Texas Tech University, Box 41051, Lubbock, TX 79409-1051, USA}

\author{Keith Bechtol\altaffilmark{4}}

\author{Jeffrey L. Carlin\altaffilmark{4}}

\author{Jay Strader\altaffilmark{7}}
\altaffiltext{7}{Department of Physics and Astronomy, Michigan State University, East Lansing, MI 48824, USA}

\author{Hu Zou\altaffilmark{8}}
\altaffiltext{8}{Key Laboratory of Optical Astronomy, National Astronomical Observatories, Chinese Academy of Sciences, Beijing 100012, China}

\author{Xu Zhou\altaffilmark{8}}
\author{Jundan Nie\altaffilmark{8}}
\author{Tianmeng Zhang\altaffilmark{8}}
\author{Zhimin Zhou\altaffilmark{8}}
\author{Xiyan Peng\altaffilmark{8}}

\begin{abstract}
We investigate the hypothesized tidal disruption of the Hercules ultra-faint dwarf galaxy (UFD). Previous tidal disruption studies of the Hercules UFD have been hindered by the high degree of foreground contamination in the direction of the dwarf.  We bypass this issue by using RR Lyrae stars, which are standard candles with a very low field-volume density at the distance of Hercules. We use wide-field imaging from the Dark Energy Camera on CTIO to identify candidate RR Lyrae stars, supplemented with observations taken in coordination with the Beijing-Arizona Sky Survey on the Bok Telescope. Combining color, magnitude, and light-curve information, we identify three new RR Lyrae stars associated with Hercules. All three of these new RR Lyrae stars lie outside its published tidal radius.  When considered with the nine RR Lyrae stars already known within the tidal radius, these results suggest that a substantial fraction of Hercules' stellar content has been stripped. With this degree of tidal disruption, Hercules is an interesting case between a visibly disrupted dwarf (such as the Sagittarius dwarf spheroidal galaxy) and one in dynamic equilibrium. The degree of disruption also shows that we must be more careful with the ways we determine object membership when estimating dwarf masses in the future. One of the three discovered RR Lyrae stars sits along the minor axis of Hercules, but over two tidal radii away. This type of debris is consistent with recent models that suggest Hercules' orbit is aligned with its minor axis.
\end{abstract}
\maketitle

\section{Introduction}
A central question in modern astrophysics is how the structures we observe today (galaxies, clusters, etc.) were formed. The $\Lambda$ cold dark matter ($\Lambda$CDM) model predicts that structure forms hierarchically. Dark matter only simulations using the $\Lambda$CDM model accurately reproduce many characteristics of our universe, but exhibit more low-mass satellite halos than we have observed as luminous satellite galaxies \citep{Kauffmann1993,Klypin1999,Moore1999}. One explanation for this discrepancy is that the majority of low-mass dark matter halos may not contain stellar matter -- they are dark. Simulations incorporating baryonic effects that suppress star formation in these low-mass halos have been successful in supporting such a model \citep[e.g.,][]{Bullock2000,Somerville2002,Koposov2009,Li2010,Brooks2013,Sawala2016a,Wetzel2016}, but it's also likely there are more luminous, low-mass satellites of the Milky Way we haven't yet discovered. To this end, the class of ultra-faint dwarf galaxies (UFDs) provide a window to find dark matter halos previously unaccounted for. \par

First discovered in the Sloan Digital Sky Survey (SDSS), UFDs are distinct from  ``classical'' dwarf galaxies in a few important ways: UFDs are more metal-poor ([Fe/H] $\leq -2$) and dark matter dominated (M/L $\sim 140 - 1700$), but less luminous, with absolute magnitudes $\text{M}_{\text{v}} \gtrsim -6$ \citep{McConnachie2012}. UFDs are also strong targets for the observation of predicted $\gamma$-ray flux from annihilation of dark matter particle candidates because of their high mass-to-light ratios \citep[e.g.,][]{Strigari2007,Bonnivard2015,Drlica-Wagner2015,Ruchayskiy2016,Moline2017}.\par

However, the high mass-to-light ratios of the UFDs have been found assuming that the radial velocities of stars in the dwarfs accurately reflect their gravitational potentials. This may not be the case if the dwarf is being tidally disrupted and unbound stars are being included in dynamical analyses \citep[e.g.,][]{Penarrubia2008,Bonnivard2015a}. Any errors in mass calculation will make comparison between the Milky Way's dwarf population and halo mass in numerical simulations inexact, and estimations of $\gamma$-ray flux from dark matter annihilation in these halos incorrect, thus thwarting two important methods for cosmological research. It is therefore critical to identify if UFDs are experiencing tidal stripping and account for it in mass determinations.\par

There is observational evidence indicating that several UFDs may be tidally disrupting. Segue 2 appears to be low luminosity for its [Fe/H] compared to the positive trend exhibited by other dwarfs of the Milky Way, M31, and dwarf irregulars of the Local Group \citep{Kirby2013}, suggesting it may have experienced tidal stripping. Bo\"{o}tes I \citep{Roderick2016a}, Canes Venatici II \citep{Sand2012}, Leo V \citep{Belokurov2008,Walker2009,Sand2012}, and Ursa Major II \citep{Munoz2010} have been found to have extended stellar features, indicating past tidal disruption. Few wide-field imaging searches sufficiently deep to detect the low surface-brightness extensions typical of tidal stripping have been performed on post-SDSS UFDs. However, most of the photometric UFD tidal debris searches \citep[e.g.,][]{Coleman2007,Sand2009,Sand2012,Roderick2015,Roderick2016a} have found some degree of extended structure, hinting that tidal disruption may be quite common for UFDs.\par

The Hercules dwarf is a prime UFD to study for tidal disruption, with high ellipticity ($\epsilon = 0.67 \pm 0.03$, \citealt{Sand2009}) and evidence for a velocity gradient across its face \citep{Aden2009}. \cite{Deason2012} found likely blue horizontal branch members up to 500 pc from the center of Hercules and further evidence for a velocity gradient consistent with \cite{Aden2009}. Several photometric studies \citep[e.g.,][]{Coleman2007,Sand2009,Roderick2015} have found Hercules to have an extended morphology, with features up to 2 kpc from the dwarf's center. These are all excellent hints that Hercules could be in the throes of tidal disruption. However, significant foreground contamination combined with the low surface brightness of Hercules limits attempts to study its extended structure. In this study, we overcome issues of foreground contamination by using RR Lyrae stars as tracers of the underlying tidal disruption of the system as a whole. \par

RR Lyrae stars are pulsating variable stars found in old, metal-poor populations that can be used as standard candles. RR Lyrae stars have been used to trace structure in the Galactic halo \citep[e.g.,][]{Sesar2010,Sesar2013} and it has been theorized that searches for overdensities of RR Lyrae stars in surveys could discover new UFDs \citep{Baker2015}. RR Lyrae stars can be well-parameterized via Fourier analysis \citep[e.g.,][]{Soszynski2008,Torrealba2015} or fitting of empirical light curve templates \citep[e.g.,][]{Sesar2010,Vanderplas2015} and differentiated from other variable stars by their colors and light curve shapes. Once characterized, empirical expressions for the intrinsic brightness of RR Lyrae stars \citep[e.g.,][]{Cacciari2003,Caceres2008} make converting an RR Lyrae star's apparent brightness into its distance simple.\par

Our approach in this paper is to use RR Lyrae stars as tracers for the tidal disruption of the Hercules UFD. Nine RR Lyrae stars in Hercules were previously found by \cite{Musella2012}. We extend this work with wide-field, time series imaging obtained with the Dark Energy Camera (DECam) and 90Prime that extends 2 kpc in projected radial distance (nine half-light radii) from the center of Hercules. Using this method, we can find individual stars with high membership likelihood farther from the body of the dwarf than any study to date, allowing us to conduct a wide and precise search for Hercules tidal debris. \par

The paper is organized as follows. In \S \ref{section:data} we describe the data and reduction procedures used in the study; in \S \ref{section:selection} we summarize the process used to select RR Lyrae stars candidates from the time series photometric catalogs; in \S \ref{section:newcandidates} we report and characterize three new RR Lyrae stars found in our field consistent with the distance of Hercules; and in \S \ref{section:tidaldestruction} we comment on the spatial distribution of these new RR Lyrae stars and draw conclusions about the degree of tidal destruction Hercules is currently exhibiting.

\begin{table}[ht]
  \centering
  \caption{Properties of Hercules}
  \begin{tabular}{c c c}
    \hline
    \hline
    Parameter & Value & Reference\\
    \hline
    R.A. & 16\textsuperscript{h}31\textsuperscript{m}03\textsuperscript{s} & \cite{Sand2009} \\
    Decl. & 12\textdegree$47^{\prime}14^{\prime\prime}$ & \cite{Sand2009} \\
    r\textsubscript{\emph{half}} & $5.91^{\prime} \pm 0.50^{\prime}$ & \cite{Sand2009} \\
    $\epsilon$ & $0.67 \pm 0.03$ & \cite{Sand2009} \\
    $\theta$ & -72.4\textdegree & \cite{Sand2009} \\
    d & $137 \pm 11$ kpc\footnote{Distance estimates range from $132 \pm 6$ kpc to $147^{+8}_{-7}$ kpc \citep[e.g.,][]{Belokurov2007,Coleman2007,Aden2009a,Sand2009,Musella2012}. This study finds 137 kpc as the most likely distance for Hercules.} & This work \\
    r\textsubscript{\emph{tidal}} & 485 pc & \cite{Aden2009} \\
    \text{[Fe/H]} & -2.41 \text{dex} & \cite{Kirby2011} \\
    \hline
  \end{tabular}
  \label{Table:assumptions}
\end{table}

\section{Observations and Data Reduction} \label{section:data}
We used wide-field imaging from DECam on the Blanco 4m telescope at Cerro Tololo Inter-American Observatory (CTIO) and the 90Prime imager on Steward Observatory's 2.3m Bok Telescope. Our DECam data included newly acquired observations, as well as archival data from \cite{Roderick2015}.\par

We imaged the Hercules dwarf in \emph{g} and \emph{r} bands using DECam. The DECam is a mosaic camera consisting of sixty-two 2K x 4K pixel CCDs, sixty of which were operational for our observations. The DECam has a scale of $0.27^{\prime\prime}$ per pixel, with a field of view of 3 square degrees. This field extends 2 kpc in projected radial distance (nine half-light radii) from the center of Hercules. \par

On 2013 March 21 and 22 we took 66 images in gray conditions, alternating exposures between \emph{g} and \emph{r} bands for a total of 33 in each. Each night we observed for two hours straight, with a median observing cadence of 3.5 minutes between each exposure. Our DECam pointings were centered $34^{\prime\prime}$ north of Hercules (see Table \ref{Table:assumptions}). Our images were not dithered. The seeing ranged from $0.9^{\prime\prime}$ to $1.6^{\prime\prime}$ with a median seeing of $1.2^{\prime\prime}$. The exposure times for all images were 180 seconds. We also obtained archival data via the NOAO Science Archive\footnote{http://archive.noao.edu/} originally published in \cite{Roderick2015} for this field consisting of 8 images of exposure length 900s in \emph{g}, with approximately the same seeing and pointing, taken with the same telescope and camera setup on 2013 July 12 and 13.\par

Normal image reductions were performed by the DECam community pipeline \citep{Valdes2014}. We additionally used masks produced by the community pipeline for each image to eliminate saturated and non-linear pixels, cosmic rays, and bleed effects.\par

To obtain more complete phase coverage, 50 images in each of \emph{g} and \emph{r} bands were also obtained in coordination with the Beijing-Arizona Sky Survey \citep[BASS;][]{Zou2017a} on the 2.3m Steward Observatory Bok Telescope with the 90Prime imager using 300s exposures. 90Prime has four 4K x 4K CCDs, each read out through four independent amplifiers, at a scale of $0.455^{\prime\prime}$ per pixel, with a field of view of 1 square degree \citep{Williams2004}. Observations were obtained between 2016 January 17 and 2016 June 5, making the timespan covered by our data 2.9 years. Images were taken in conditions that were non-optimal for the survey -- as such, the quality of the Bok images was inconsistent, with seeing from $1.5^{\prime\prime}$ to $3^{\prime\prime}$ and varying degrees of sky background, requiring careful attention to extract accurate photometry. Bias subtraction and flat-fielding was performed using normal routines from PyRAF\footnote{http://www.stsci.edu/institute/software\_hardware/pyraf} and astrometric solutions were corrected using a local installation of the \texttt{astrometry.net} code base \citep{Lang2010}. Our Bok pointing of $\alpha=$16\textsuperscript{h}31\textsuperscript{m}10\textsuperscript{s}, $\delta=$12\textdegree$25^{\prime}01^{\prime\prime}$ was optimized for the primary candidates found in our DECam imaging (for a description of the candidate selection procedure, see \S \ref{section:selection}). Our images were not dithered.\par

\subsection{Photometry and Calibration} \label{subsection:photometry}
We performed stellar photometry on our images using the \textsc{daophot} and \textsc{allstar} packages \citep{Stetson1987}. We photometered each chip separately, using the average full-width at half-maximum (FWHM) values for each chip as input for \textsc{daophot} and 5x the FWHM as the radius to fit the point spread function (PSF). For PSF fitting we used a five-parameter Penny function and allowed it to vary quadratically over each chip. We ran two passes of \textsc{allstar}: the first on the image, and the second on the image with detections from the first pass removed, allowing us to find faint stars missed on the first pass. For our Bok images, we followed the same process, although in some cases we manually selected stars to use in constructing the PSF to produce a good fit.\par

We calibrated the photometry by bootstrapping the data onto SDSS DR12 photometry and correcting for Galactic extinction. We calibrated our instrumental magnitudes by matching our photometric catalog to SDSS DR12  \citep{Alam2015} and using the stars found in common to fit a zeropoint and colorterm for each CCD chip in each image. We selected only sources with $0 < \text{chi} < 1$ and $|\text{sharp}| < 0.1$ for the calibration, which are measures of the goodness-of-fit and the shape of the PSF, respectively. These values are returned by \textsc{allstar} and were selected to eliminate galaxies from our calibration. We additionally imposed color and magnitude restrictions of $(g-r) < 1$ and $18 < g < 20$. There are many red foreground stars in the direction of Hercules, and we found the color cut allowed us to find robust colorterms for our objects of interest. We corrected the instrumental magnitudes for Galactic extinction using the dust maps from \cite{Schlegel1998} assuming R\textsubscript{V} = 3.1 and the updated reddening coefficients for SDSS filters from \cite{Schlafly2011}.\par

We used the general form of the maximum likelihood technique from \cite{Boettcher2013} to calibrate our data to SDSS without spatial correction terms, as there are no residual trends in pixel position for the DECam or 90Prime instruments. We determined uncertainties on the zeropoints and colorterms by performing a 1000 iteration bootstrap and added these in quadrature to the photometric uncertainties returned by \textsc{allstar}. RR Lyrae stars at the distance of Hercules have $g_0 \approx 21.1$ and $r_0 \approx 21.0$, corresponding to errors of 0.075 magnitude in our DECam and Bok data for signal-to-noise of 14.49.\par

\begin{figure}[h]
	\centering
        \includegraphics[width=0.5\textwidth,page=2]{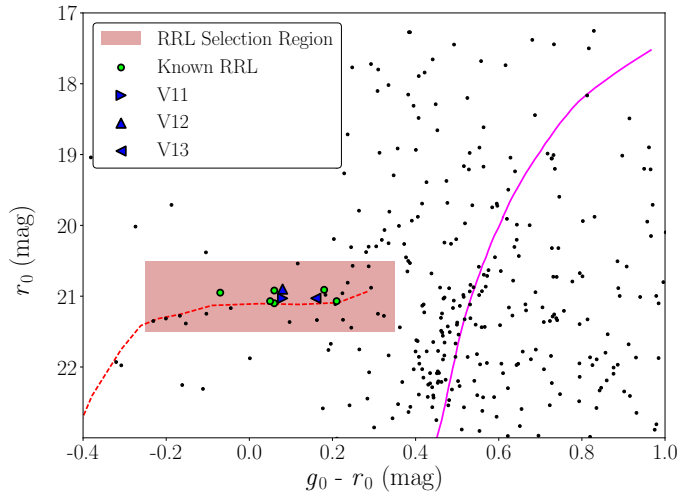}
	\caption{Color-magnitude diagram of all sources within the half-light radius of Hercules. The RR Lyrae star candidate selection region used in this paper, the \cite{Musella2012} RR Lyrae stars found in our DECam discovery data, and our new RR Lyrae stars are shown. Magnitudes for each source are medians over the DECam and 90Prime time series observations and corrected for Galactic extinction (see Section \ref{subsection:photometry}). Overplotted is a horizontal branch constructed from the M15 photometry of \cite{An2008}, shifted to distance modulus $\mu_0 = 20.68$, or a distance of 137 kpc, to match the horizontal branch apparent magnitude of Hercules, with a 12.7 Gyr, $\text{Z}=0.0001$ isochrone at the same distance from CMD 2.7 using PARSEC 1.2S \citep{Aringer2009,Bressan2012,Chen2014}.}
	\label{cmd}
\end{figure}

\section{RR Lyrae Star Candidate Selection} \label{section:selection}
All RR Lyrae star candidate selection was based on our DECam data, due to the larger field of view -- our Bok pointing was then optimized for the most promising of the primary candidates from the DECam imaging. \par

To select RR Lyrae star candidates we begin by implementing selection criteria 20.5 $\leq \hspace{1mm} \emph{r}_0 \hspace{1mm} \leq$ 21.5, overall change in magnitude of at least 0.2 in \emph{g} and \emph{r}, and color $-0.25 \leq (g_0-r_0) \leq 0.35$. The magnitude range will allow us to find RR Lyrae stars in our field of view with distances within $\pm$50 kpc of Hercules (see \S \ref{section:tidaldestruction}), and the color and overall change in magnitude criteria were based on the properties of RR Lyrae stars found in the SDSS Stripe 82 survey \citep{Sesar2010} and the La Silla-QUEST survey \citep{Zinn2014}. Figure \ref{cmd} shows the CMD of sources within the half-light radius of Hercules. Previously known RR Lyrae stars from \cite{Musella2012} are shown with circles and the shaded box highlights the color-magnitude space explored in this paper for new RR Lyrae stars. Magnitudes for each source are medians over the DECam and 90Prime time series observations. We overplot a horizontal branch generated from the \textsc{daophot}/\textsc{allstar} photometry of \cite{An2008} using SDSS imaging, and a 12.7 Gyr, $\text{Z}=0.0001$ isochrone from CMD 2.7 using PARSEC 1.2S \citep{Aringer2009,Bressan2012,Chen2014}. We corrected the horizontal branch for reddening with E(B-V)=0.11 \citep{Schlegel1998}, R\textsubscript{V} = 3.1, and the updated reddening coefficients for SDSS filters from \cite{Schlafly2011}. The metallicity ([Fe/H]=$-$2.3, \citealt{Carretta2009}) and age \citep[12.75 Gyr,][]{VanderBerg2013} of M15 are similar to Hercules, and M15 also exhibits a prominent horizontal branch we can use to calculate the distance of Hercules. We use a shift of +5.57 \emph{g} mag to match the horizontal branch fiducial of M15 to the horizontal branch of Hercules. Assuming the reddening-corrected distance modulus of M15 to be 15.11 \citep{VandenBerg2016}, we have a total distance modulus of $\mu_0=20.68$ for a distance of 137 kpc, consistent with literature values for Hercules \citep[e.g.,][]{Belokurov2007,Coleman2007,Sand2009,Musella2012} and our distance calculation in Section \ref{subsection:demographics} using RR Lyrae stars as standard candles. Table \ref{Table:assumptions} summarizes the Hercules parameters assumed here, and throughout the remainder of this paper.\par

\begin{figure*}[t!]
  \includegraphics[width=.5\linewidth, page=2]{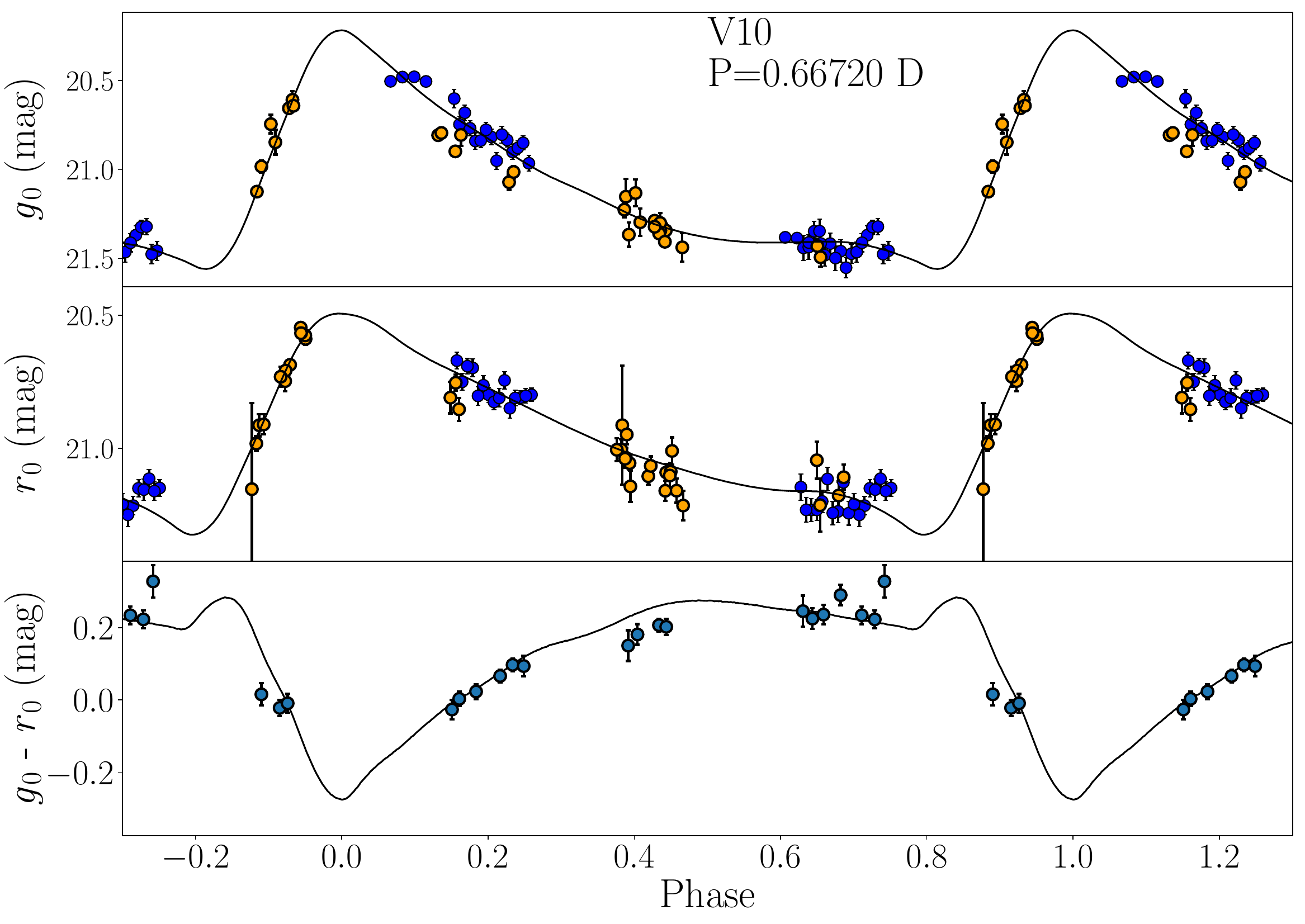}
  \includegraphics[width=0.5\linewidth, page=62]{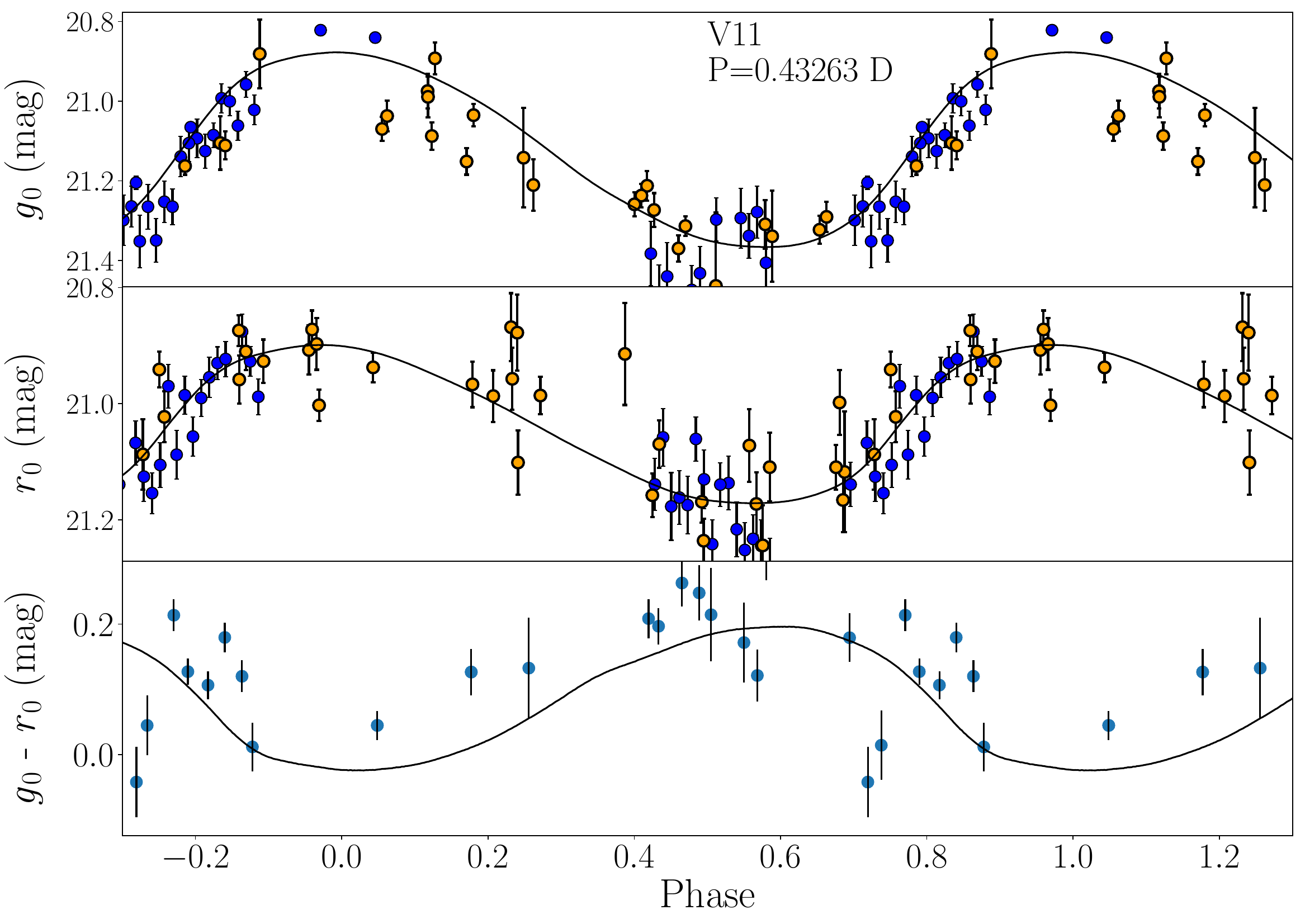}
  \includegraphics[width=0.5\linewidth, page=50]{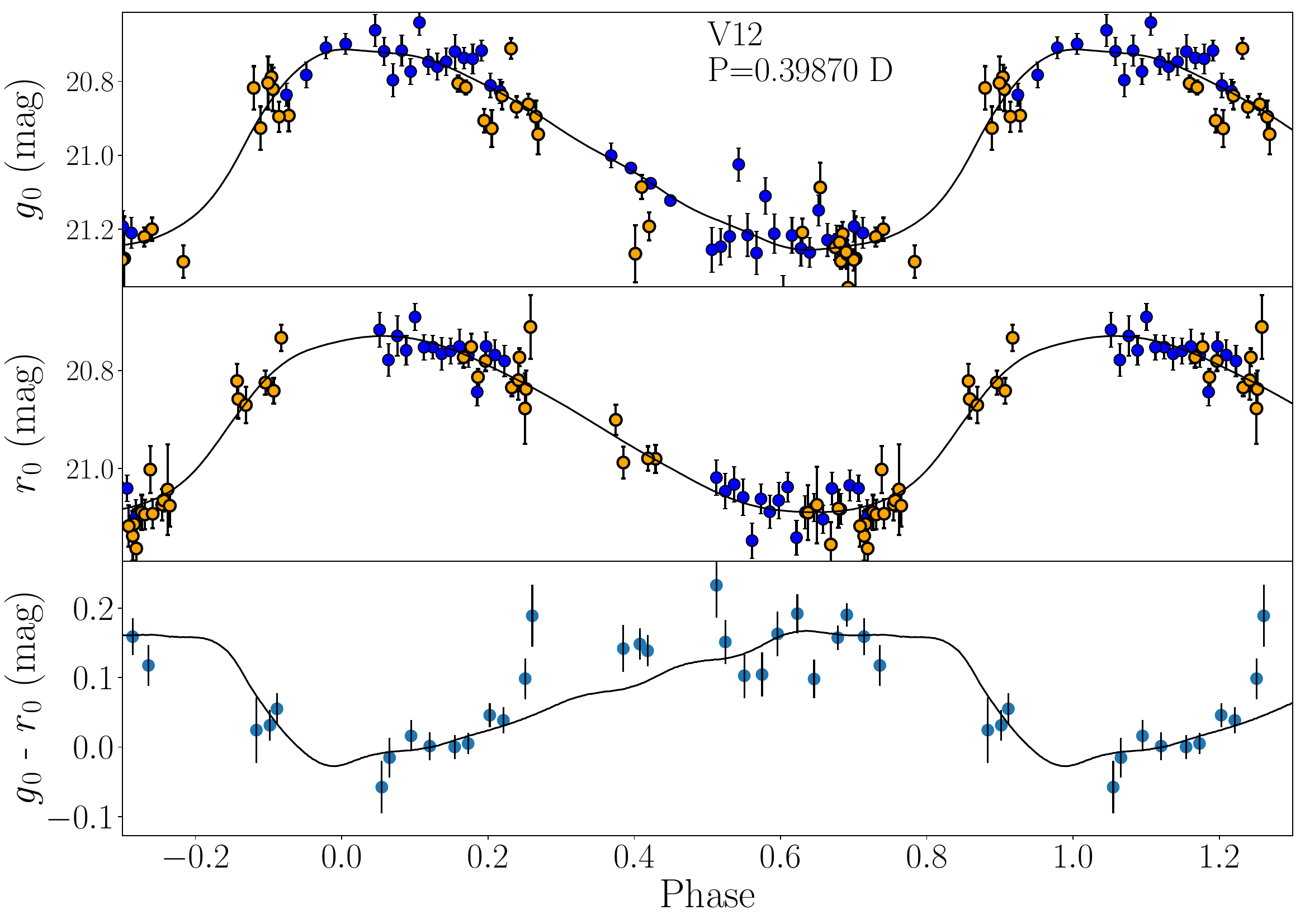}
  \includegraphics[width=0.5\linewidth, page=56]{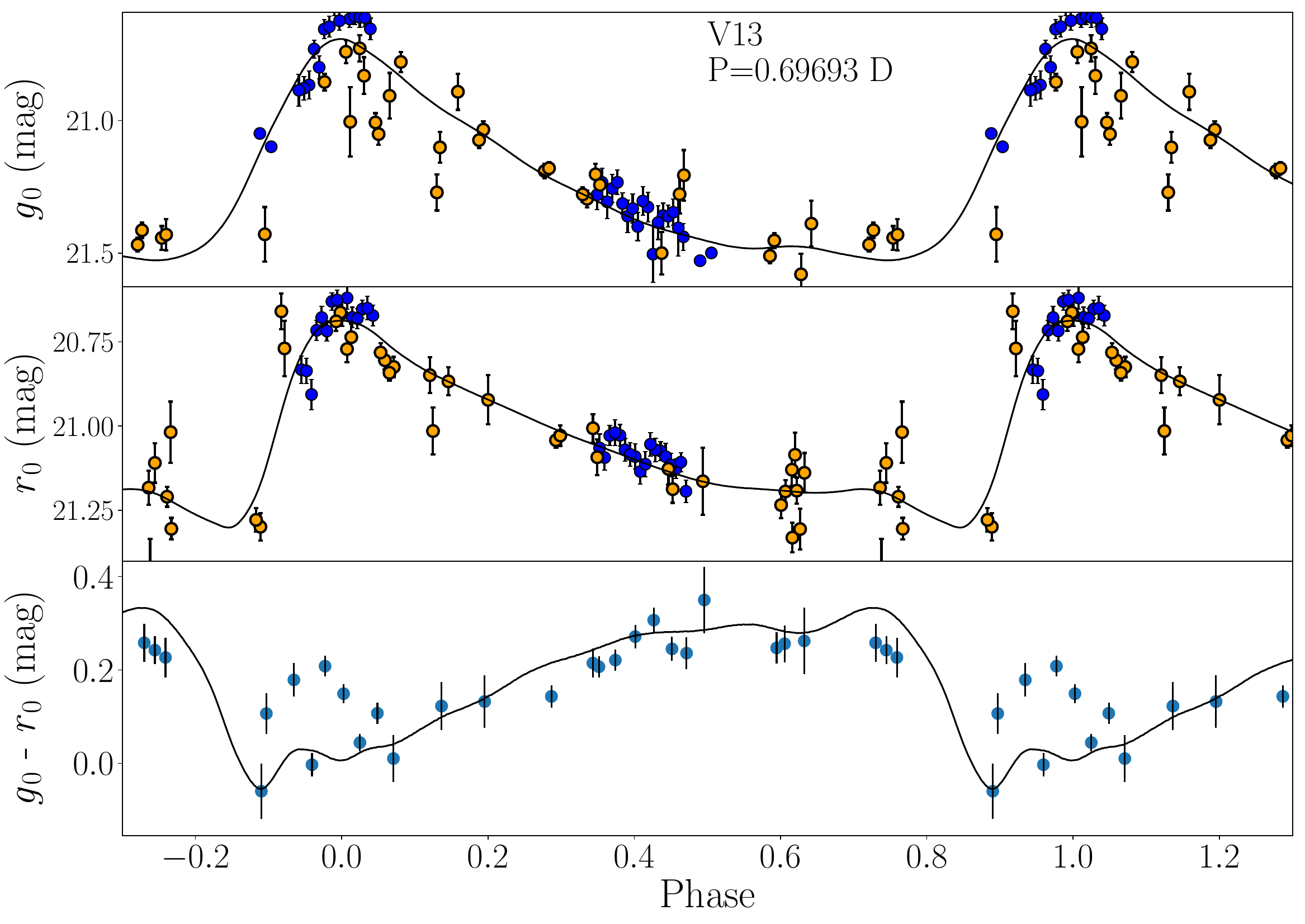}
  \caption{Period-folded light curves for the RRab V10 from \cite{Musella2012} and the three new RR Lyrae stars from this work. We used the multi-band periodogram described in \cite{Vanderplas2015} coupled with fits to the templates from \cite{Sesar2010} to characterize the periods, amplitudes, and flux-averaged magnitudes of our variables. Blue data points are DECam observations and orange points are Bok observations. To generate the \emph{g}\textsubscript{0}-\emph{r}\textsubscript{0} data, we bin our observations such that each point on the light curve is derived from no less than two observations with phase separation no greater than 0.05.}
  \label{2lightcurves}
\end{figure*}

To quantify variability, we use the approach from \cite{Boettcher2013} where we characterize the difference between the error-weighted, sigma-clipped average magnitude of a star and the \emph{i}th exposure to be
\begin{equation} \label{vareq}
\delta_{mag, i}=\left| \frac{\langle m \rangle - m_{i}}{\sqrt{\sigma^{2}_{\langle m \rangle} + \sigma^{2}_{i}}} \right| 
\end{equation}
To be selected as a candidate RR Lyrae star, we require $\delta_{mag, i} > 2.0$ for at least three exposures in both \emph{g} and \emph{r} bands in our DECam imaging.\par

This selection left us with $\sim$100 possible RR Lyrae star candidates over the full DECam field of view. Garling, Willman, Sand, and Hargis visually inspected the raw light curves of these candidates to determine which were most likely to be RR Lyrae stars. Visual inspection of these objects showed many of them had amplitudes much greater than expected for RR Lyrae stars, possibly because they are eclipsing binary systems or because of abnormally high photometric errors. \par

From these $\sim$100 candidates, we selected thirteen primary candidates and twenty-seven secondary candidates. The primary candidates we selected to analyze were detected in more than twenty images in both \emph{g} and \emph{r} bands in our DECam images and had distinct time variation in their raw light curves with visually estimated amplitudes and periods similar to RR Lyrae stars from \cite{Sesar2010} and \cite{Musella2012}. The secondary candidates either had less than twenty detections in \emph{g} or \emph{r}, higher or lower amplitudes than expected, or varied on time scales that seemed too short or too long.\par

We found that seven of our primary candidates were known RR Lyrae stars and one was an Anomalous Cepheid (AC) published by \cite{Musella2012}  -- the other two of their published RR Lyrae stars fell in chip gaps in our DECam discovery images. Thus out of our original 40 candidates, 32 were new. The fact that we independently selected the variables of \cite{Musella2012} as primary candidates is evidence that the criteria used in the visual inspection were reasonable.\par

Following the recommendation of \cite{Vanderplas2015}, we used a two-step process to characterize the periods of the known variables and our 32 remaining RR Lyrae star candidates. We begin by using the multi-band Lomb-Scargle (LS) periodogram from \cite{Vanderplas2015} as implemented in ``General Tools for Astronomical Time Series in Python'' (GATSPY)\footnote{http://www.astroml.org/gatspy/}. For the periodogram, we used a shared-phase, single-term Fourier model, as it has been shown that the light curves of RR Lyrae stars are synchronized in phase across filters \citep[e.g.,][]{Sesar2010}, so that extrema of an RR Lyrae star's light curves in different filters occur at the same time. This process typically output a periodogram with several peaks of similar power. To pick out the best period among these possibilities, we fit templates from \cite{Sesar2010} to our data at periods corresponding to the top five peaks in the multi-band periodogram.\par

In our data, this two-step process is generally successful in overcoming what \cite{Sesar2017} calls the ``non-physical'' nature of the multi-band periodogram. We find that the greatest peak in the multi-band periodogram identifies the correct period for only three of the eight RR Lyrae stars in common between this work and \cite{Musella2012}, but using template fitting to discriminate between the top five periods from the multi-band periodogram reproduces seven of the eight periods as the most statistically significant result. We therefore find this combined approach to be an excellent compromise between highly accurate but computationally-intensive template-fitting and the fast but ``non-physical'' multi-band periodogram.

For the single RR Lyrae star where the period from \cite{Musella2012} was not reproduced by our two-step process process, V8, their period produced the \emph{second} best fit, with relative goodness-of-fit statistics $< 3$\% different from the top period found by the numerical optimization. In this situation, physical information from the phase-color relationship can typically break the numerical degeneracy in the goodness-of-fit statistic. RR Lyrae stars are bluest at maximum light, grow redder until a peak halfway through their period, then proceed back to their bluest at maximum light. For V8, this was clearly seen in the second best period (the period reported by \citealt{Musella2012}) and not in the top period reported by the template fitting routine. We also note that our light curve for V8 has far worse phase coverage than any other variable we analyzed (see Appendix \ref{appendix:lightcurves}), making it understandable why the routine did not reproduce the period from \cite{Musella2012} on the first try.\par

To differentiate between real RR Lyrae stars and contaminate variables, we combined goodness-of-fit, color-phase, and period-amplitude information. Based on a visual comparison between these properties and those expected for RR Lyrae stars, we identify three new RR Lyrae stars (V11, V12, and V13) out of 32 candidates. Our naming scheme follows \cite{Musella2012}, by increasing distance from Hercules. We conducted the visual inspection described above to select the best periods for these RR Lyrae stars, but found that the top period from the template fitting procedure was the optimal period for all three of the new RR Lyrae stars. Light curves for these new RR Lyrae stars and V10 from \cite{Musella2012} are shown in Figure \ref{2lightcurves}. We also discover a variable with a period of $\sim$1 day ($\alpha=16\textsuperscript{h}32\textsuperscript{m}20.61\textsuperscript{s}, \delta=12$\textdegree$40^{\prime}09.95^{\prime\prime}$), but due to poor phase coverage of its light curve, we are unable to constrain its other properties. With such a period, it is unlikely to be an RR Lyrae star, though about 4\% of RR Lyrae stars have periods greater than 1 day \citep{Torrealba2015,Drake2017}, so an RR Lyrae star classification is not ruled out. However, due to our insufficient phase coverage ($<30$\%, and only during the rise) and a poor fit to the RR Lyrae star templates from \cite{Sesar2010}, we exclude this candidate from the remainder of our analysis and discussion because we cannot be sure that our characterization of this star is robust. One of our new RR Lyrae stars, V11, falls into the field of view of \cite{Musella2012}, but was excluded from their final results on the basis of a scattered light curve and low amplitude (Musella 2015, private communication).\par

\begin{table*}[t]
\caption{Properties of Variable Stars in Hercules}
\begin{tabular}{c c c c c c c c c c c} 
\hline
\hline
ID & RA & DEC & Type & $\langle g_0 \rangle$ & $\langle r_0 \rangle$ & $\langle g_0 - r_0 \rangle$ & $\text{A}_{g}$ & $\text{A}_{r}$ & Period (Musella et. al 2012) & Period \\
 & (hms) & (dms) & & (mag) & (mag) & (mag) & (mag) & (mag) & (days) & (days)\\
\hline
V1\tablenotemark{a} & 16:21:02.17 & +12:47:33.7 & RRab & - & - & - & - & - & 0.639206 & - \\
V2 & 16:31:02.87 & +12:45:48.47 & AC & 20.60 & 20.49 & 0.11 & 0.53 & 0.33 & 0.53777 & 0.53778  \\
V3 & 16:30:54.89 & +12:47:04.23 & RRc & 21.16 & 21.08 & 0.08 & 0.55 & 0.41 & 0.39997 & 0.40853 \\
V4 & 16:30:56.09 & +12:48:29.25 & RRc & 21.10 & 21.04 & 0.06 & 0.63 & 0.44 & 0.39576 & 0.39576 \\
V5 & 16:30:52.24 & +12:49:12.03 & RRc & 21.12 & 21.07 & 0.05 & 0.54 & 0.35 & 0.40183 & 0.40182 \\
V6 & 16:30:52.37 & +12:50:00.03 & RRab & 21.28 & 21.07 & 0.21 & 0.85 & 0.63 & 0.69981 & 0.69976 \\
V7\tablenotemark{a} & 16:31:29.48 & +12:47:34.9 & RRab & - & - & - & - & - & 0.67799 & - \\
V8\tablenotemark{b} & 16:31:27.11 & +12:44:16.74 & RRab & 20.88 & 20.95 & $-$0.07 & 1.6 & 0.83 & 0.66234 & 0.66619 \\
V9 & 16:31:29.43 & +12:40:03.35 & RRab & 21.09 & 20.91 & 0.18 & 0.89 & 0.64 & 0.72939 & 0.72940 \\
V10 & 16:30:04.03 & +12:52:05.95 & RRab & 20.98 & 20.92 & 0.06 & 1.3 & 0.83 & 0.6616 & 0.66720\\
\hline
V11 & 16:30:27.15 & +12:55:41.34 & RRc & 21.11 & 21.03 & 0.08 & 0.49 & 0.27 & - & 0.43263\\
V12 & 16:30:55.81 & +12:27:09.21 & RRc & 20.98 & 20.90 & 0.08 & 0.54 & 0.36 & - & 0.39870\\
V13 & 16:32:27.11 & +12:43:28.03 & RRab & 21.19 & 21.03 & 0.16 & 0.83 & 0.61 & - & 0.69693\\

\hline
\end{tabular}
\tablenotetext{a}{These stars fell in chip gaps in our DECam discovery data.}
\tablenotetext{b}{Our light curve for V8 has poor phase coverage, so mean magnitudes and amplitudes may not be robust.}
\label{Table:properties}
\end{table*}

\section{New RR Lyrae Stars in Hercules} \label{section:newcandidates}
We present three new RR Lyrae stars in our field of view and the results of our analysis for the variables found by \cite{Musella2012} in Table \ref{Table:properties}. We computed flux-averaged magnitudes by converting the best-fit template models from magnitudes to flux counts\footnote{http://www.sdss.org/dr12/algorithms/fluxcal/}, taking integral averages, and converting back to magnitudes. The amplitudes and flux-averaged magnitudes reported are computed from the best-fit templates. Period-folded light curves using our DECam and 90Prime data are shown for one of the RR Lyrae stars from \cite{Musella2012} and our new RR Lyrae stars in Figure \ref{2lightcurves}. Additional light curves for RR Lyrae stars published in \cite{Musella2012} can be found in Appendix \ref{appendix:lightcurves}. DECam observations are in blue and Bok observations are in orange. To generate the \emph{g}\textsubscript{0}-\emph{r}\textsubscript{0} data, we bin our observations such that each point on the light curve is derived from no less than two observations with phase separation no greater than 0.05.\par

\subsection{The RR Lyrae Star Demographics of Hercules} \label{subsection:demographics}
\begin{figure}[!ht]
  	\centering
  	\includegraphics[width=0.5\textwidth, page=68]{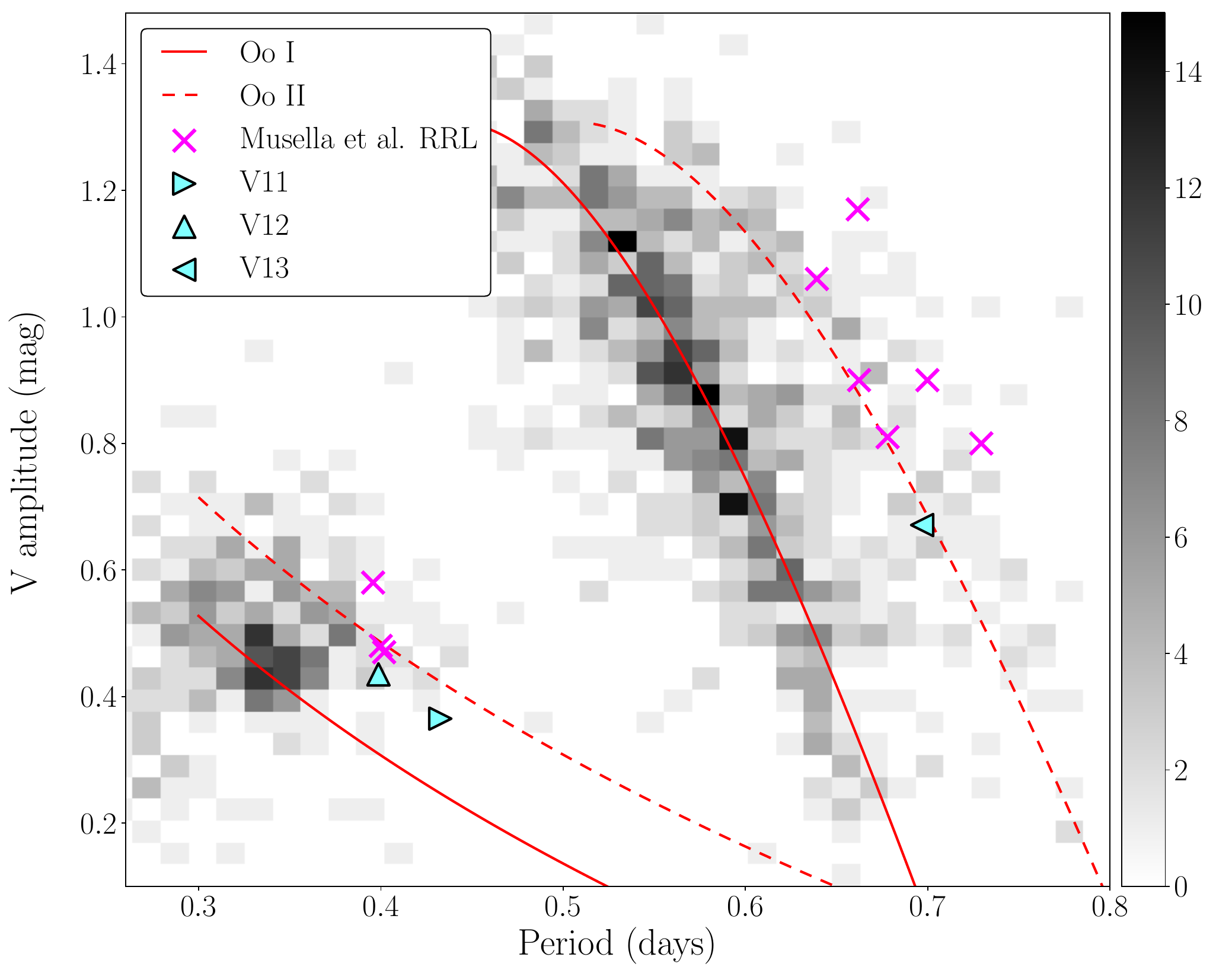}
	\caption{Period-amplitude (Bailey) diagram showing the RR Lyrae stars from \cite{Musella2012} and this work, overlaid onto the amplitudes and periods of RR Lyrae stars from the La Silla-QUEST survey \cite{Zinn2014}. For our new RR Lyrae stars, we convert the best fit light curve templates to V band using the filter conversions from \cite{Jordi2006} and calculate their amplitudes. For the RR Lyrae stars discovered in \cite{Musella2012}, we adopt their amplitudes. The parameterizations of the period-amplitude relationship for RRab and RRc in Oosterhoff I and II populations from \cite{Zorotovic2010} are overlaid as solid and dashed lines, respectively.}
	\label{bailey_diagram}
\end{figure}

This section will discuss the results of our RR Lyrae star search in the context of the RR Lyrae stars found by \cite{Musella2012} and will also compare the updated RR Lyrae star population of Hercules to those of other dwarf galaxies. We begin by constructing the period-amplitude diagram (also known as the Bailey diagram) for the full sample of Hercules RR Lyrae stars in Figure \ref{bailey_diagram}. To compute V band amplitudes for our new RR Lyrae stars, we convert our best fit light curve templates in \emph{g} and \emph{r} to V using the filter conversions from \cite{Jordi2006}. We adopt the amplitudes from \cite{Musella2012} for the variables characterized in that work. RR Lyrae stars from the La Silla-QUEST survey \citep{Zinn2014} are presented as a histogram for comparison. The relationships between period and amplitude for Oosterhoff I and II populations\footnote{First discovered in the Milky Way's globular clusters, Oosterhoff I and II objects have mean RRab periods of 0.55 days and 0.65 days, respectively \citep{Oosterhoff1939}. For a review of modern developments, see \cite{Catelan2009}.} from \cite{Zorotovic2010} are overplotted as solid and dashed lines, respectively. The RR Lyrae stars of Hercules align well with the Oosterhoff II lines.\par

\begin{figure*}[t]
  \centering
  \includegraphics[width=0.75\linewidth, page=70]{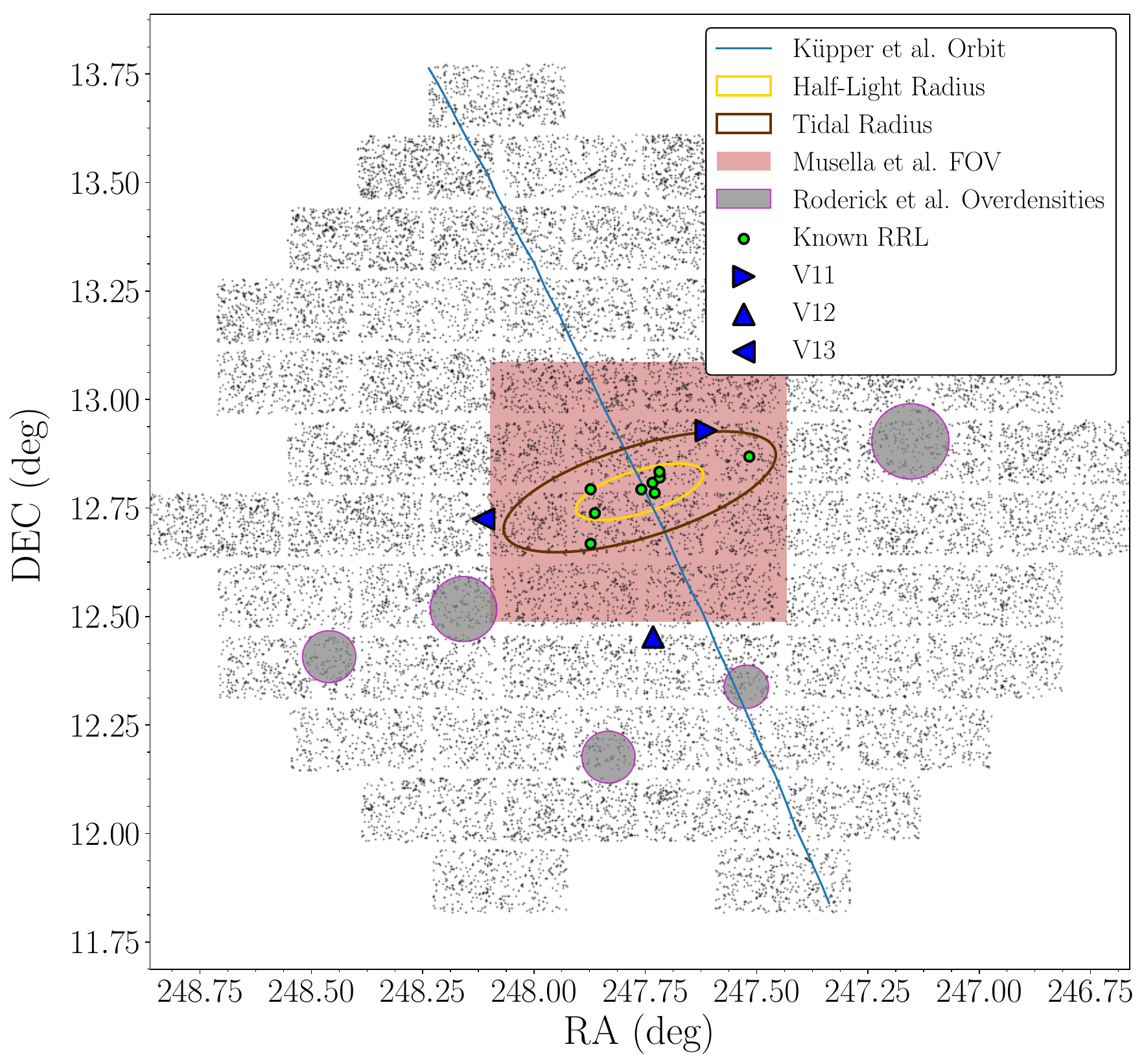}
  \caption{The spatial distribution of our new RR Lyrae stars (differently oriented blue triangles), with the field of view (shaded box) and RR Lyrae stars from \cite[][green points]{Musella2012}. The half-light radius \cite[$r_{h} = 5.91^{\prime}\pm0.50^{\prime}$,][gold ellipse]{Sand2009}, tidal radius \cite[485 pc,][brown ellipse]{Aden2009}, and overdensities from \cite[][grey shaded circles]{Roderick2015} are also shown. The location of V12 along the minor axis of the dwarf is consistent with the models of \cite{Kupper2017}, which suggest that the minor axis of Hercules is aligned with its orbit and predict some tidal debris along this path.}
  \label{distributionplot}
\end{figure*}

With $\langle \text{P}_{ab} \rangle = 0.68 \pm 0.03$ days, Hercules exhibits the properties of an Oosterhoff II system, like several other UFDs 
\citep[for a review of the Oosterhoff types of UFDs, see][]{Clementini2014}.The RRab discovered in our study (V13) has a period of 0.69693 days -- consistent with the periods of the known RRab in Hercules. V12 is an RRc with period 0.39870 days, consistent with the mean period of $\langle \text{P}_{c} \rangle =$ 0.40 days found for the three previously known RRc variables in Hercules. V11 has a longer period at 0.43263 days -- this is about 35 minutes longer than V3 from \cite{Musella2012}, the RRc with the longest period in their analysis.\par

We now determine the distance to Hercules from the apparent magnitude of its RR Lyrae stars. We begin by using the metallicity of Hercules to calculate the absolute magnitude in the V band expected for its RR Lyrae stars using the slope and zeropoint from \cite{Cacciari2003},
\begin{equation}
  \text{M}_{\text{V}} = \left (0.23 \pm 0.04 \right) \ [\text{Fe}/\text{H}] + \left( 0.93 \pm 0.12 \right)
\end{equation}
and find $\text{M}_{\text{V}} = 0.38 \pm 0.16$ mag with [Fe/H] = -2.41 dex and $\sigma_{\text{[Fe/H]}} = 0.56$ \citep{Kirby2011}. Next we convert the mean magnitudes of the RR Lyrae stars, $\langle g_0 \rangle = 21.09 \pm 0.11$ and $\langle r_0 \rangle = 21.00 \pm 0.09$, to V band using the relations from \cite{Jordi2006} and obtain $\text{m}_\text{V,0} = 21.06 \pm 0.07$ mag, consistent with that found by \cite{Musella2012}. We find a distance modulus of $\mu_0=20.68 \pm 0.17$ for distance $137 \pm 11$ kpc, consistent with previous distance estimates \citep[e.g.,][]{Belokurov2007,Coleman2007,Sand2009,Musella2012}.\par

\section{Tidal Destruction of Hercules} \label{section:tidaldestruction}
The spatial distribution of RR Lyrae stars in Hercules is visualized in Figure \ref{distributionplot}, with the RR Lyrae stars from \cite{Musella2012} shown as circles and our new RR Lyrae stars as triangles. We show the half-light and tidal radii of Hercules, as well as the field of view from \cite{Musella2012} and the orbit of Hercules from \cite{Kupper2017}. We plot other sources from the photometry as black dots to illustrate our full field of view.\par

Previous studies of the Hercules UFD have shown evidence of tidal debris that lie outside the main body of Hercules \citep{Coleman2007,Sand2009,Deason2012,Roderick2015}. The DECam study of \cite{Roderick2015} traced possible Hercules substructure over an area equal to our own DECam imaging and provides an excellent reference for comparison of the locations of our new RR Lyrae stars to possible Hercules tidal debris. The parameters assumed for this analysis are summarized in Table \ref{Table:assumptions}. \par

We find that our new RRc V11 falls in the central overdensity surrounding Hercules identified in \cite{Roderick2015}. The RRc V12 is $\approx 2.7^{\prime\prime}$ (115 pc) from segment 7 of \cite{Roderick2015}. This is one of the smaller overdensities identified in their work, but there are several other overdensities in this region and V12 is closer to Hercules than the overdensities from \cite{Roderick2015}, suggesting there could be debris loosely distributed from the dwarf out to these farther overdensities. Lastly, the RRab V13 appears $\approx 3^{\prime\prime}$ (128 pc) north of segment 9 in \cite{Roderick2015}; similar to the case for V12, there are a number of other overdensities detected in that region, leading us to believe tidally stripped Hercules members could be sparsely distributed in this region. \par

To assess the possibility of our new RR Lyrae stars being Milky Way halo stars instead of tidal debris from Hercules, we look at new data on the RR Lyrae star density as a function of Galactocentric distance from the High Cadence Transient Survey (HiTS; \citealt{Forster2016,Medina2017}, in preparation), which indicate a RR Lyrae star halo density of $\sim 10^{-4}$ kpc\textsuperscript{-3} at the distance of Hercules, where their completeness is $\sim$ 85\%. To estimate the volume probed by our RR Lyrae star search, we first find the area covered by DECam's field of view to be 16.7 kpc\textsuperscript{2}. To find the depth probed by our search, we convert the faintest and brightest magnitudes in our search (\emph{g}\textsubscript{0} = 21.85, \emph{r}\textsubscript{0} = 21.5 and \emph{g}\textsubscript{0} = 20.25, \emph{r}\textsubscript{0} = 20.5 respectively) to V band through the relations of \cite{Jordi2006} to find m\textsubscript{V,0} = 21.67 and 20.41 as the limits of our search. We then use M\textsubscript{v} = 0.38 as the absolute magnitude of RR Lyrae stars in Hercules, as calculated in Section \ref{subsection:demographics}, so that the depth of our search is 79.7 kpc, giving us a total volume for our RR Lyrae star search of 1331 kpc\textsuperscript{3}. Then using a RR Lyrae star halo density of $\approx 10^{-4}$ kpc\textsuperscript{-3} at the distance of Hercules \cite[][in preparation]{Medina2017}, we find that we would expect 0.13 RR Lyrae stars our in search volume. We draw 10,000 samples from a Poisson distribution with $\lambda=0.13$ and find that 12\% of our samples contain one field RR Lyrae star and only 1\% of samples contain two or more field RR Lyrae stars. If we instead use the broken power-law fit from \cite{Zinn2014} for the RR Lyrae star density profile, we find a RR Lyrae star halo density of $\sim$ 10\textsuperscript{-4.5} kpc\textsuperscript{-3} at the distance of Hercules, resulting in 4\% of our Poisson samples containing one field RR Lyrae star, and only 8 of our 10,000 samples containing more than one field RR Lyrae star.
This shows that the probability of there being a field RR Lyrae star in our search volume is low, but not insignificant, while the probability of more than one field RR Lyrae star being present is very low.\par

To use the spatial distribution of the RR Lyrae stars in Hercules to estimate the extent of its disruption, we first test the hypothesis that the RR Lyrae stars trace the underlying distribution of stars from Hercules. We perform a Kolmogorov-Smirnov (KS) test \citep{Kolmogorov1933,Smirnov1948} on the radial distribution of proper-motion-selected stars from \cite{Fabrizio2014} and the distribution of Hercules RR Lyrae stars to determine the likelihood of the two samples belonging to the same underlying population. We find D=0.181 and p=79.5\%, which shows good agreement between the distribution of proper-motion-selected members of Hercules and the RR Lyrae star population and supports our assumption that the RR Lyrae stars of Hercules trace its underlying stellar population. Therefore, with three of the twelve Hercules RR Lyrae stars outside the dwarf's tidal radius\footnote{It is worth noting this value was determined by assuming that the tidal radius lies just within the farthest members of Hercules determined by their Str{\"o}mgren photometry -- thus this determination is inherently imprecise.}, it is likely that a substantial fraction of its stellar content has been tidally stripped.\par

The spatial distribution of Hercules RR Lyrae stars is also consistent with the hypothesis that the dwarf's minor axis is aligned with its orbit. \cite{Kupper2017} used simulations to reproduce the ellipticity and position angle of Hercules, with the dwarf's minor axis aligned with its orbit. In this model, the elongation of Hercules is produced by tidal shocking from the dwarf's last pericenter passage coupled with differential orbital plane precession resulting from a non-spherical Galactic potential. This model predicts tidally stripped stars to be present along the dwarf's minor axis, similar to where we find our new RRc V12. A more traditional explanation for Hercules' elongation would be that Hercules is a dwarf galaxy in the midst of disrupting into a stellar stream, so that the dwarf's major axis is aligned with its orbital path. Studies have considered orbits that align with the major axis of Hercules \citep[e.g.,][]{Martin2010}, but such models have struggled to reproduce the observed properties of Hercules \citep[e.g.,][]{Blana2015}. Additionally, these models do not predict tidal debris along the dwarf's minor axis, so that RRc V12 does not fit into their predicted tidal debris structure.\par

The conclusion that Hercules is being tidally disrupted certainly implies that the main body of the Hercules dwarf had greater stellar mass and luminosity in the past than it does at the present day. Hercules already stands as an outlier on the metallicity-luminosity relationship for dwarfs, with $L_\text{V}/L_{\odot}$ twelve times higher than expected by the average relationship found for Milky Way and M31 dwarf spheroidals and Local Group dwarf irregulars \citep[][and references therein]{Kirby2013}. Increasing $L_\text{V}/L_{\odot}$ by 25\% as an estimate for the stellar mass stripped from Hercules does little to affect this, resulting in a change that is smaller than the uncertainty on $L_\text{V}/L_{\odot}$. Thus, tidal stripping of stars does not, on the surface, explain Hercules' place on the luminosity-metallicity relationship. However, most other UFDs have metallicities that would imply higher luminosities than they are observed to have, which could be a result of tidal disruption and an avenue for further dynamical studies of UFDs. \par

\section{Conclusion}
In this work, we use CTIO DECam and Bok 90Prime data to conduct a wide-field search for RR Lyrae stars around the Hercules UFD with the goal of constraining its tidal disruption. A previous study of RR Lyrae stars in Hercules by \cite{Musella2012} found nine RR Lyrae stars in a field of view half the size of that presented in this paper, suggesting that a wider field study could yield additional RR Lyrae stars. With DECam's 3 square degree field of view, we were able to search for RR Lyrae stars nine half-light radii from the center of the dwarf. The low field density of $\approx 10^{-4}$ RR Lyrae stars kpc\textsuperscript{-3} at the distance of Hercules \cite[][in preparation]{Medina2017} means that RR Lyrae stars at this distance are likely to be associated with the Hercules UFD. \par

We recovered seven of the nine RR Lyrae stars in \cite{Musella2012} (two others fell in chip gaps of our DECam discovery data) and also discovered one new RRab and two new RRc members of Hercules. The new RR Lyrae stars are consistent with periods and amplitudes typical of Oosterhoff II objects.\par

We find that two of our new RR Lyrae stars are distributed along the dwarf's major axis, as predicted for the majority of debris from tidally disrupting satellites, while one of our RR Lyrae stars is near the dwarf's minor axis, consistent with the tidal disruption models of \cite{Kupper2017}, which predict the bulk of tidal debris to be distributed along the observed major axis, but with additional debris along the dwarf's minor axis as a result of tidal shocking coupled with an orbital trajectory aligned parallel to the dwarf's minor axis. Attempts to model the orbit of Hercules along its major axis, under the assumption that it is disrupting into a tidal stream, do not predict such tidal debris along the minor axis \citep{Martin2010,Blana2015}. \par

The distribution of our new RR Lyrae stars, with all three outside the tidal radius, suggests that Hercules has experienced significant tidal stripping. This degree of tidal disruption can inflate mass estimates if dynamic equilibrium is assumed \citep{Penarrubia2008,Bonnivard2015a}. Biased mass estimates will impact studies of small-scale structure formation and studies that seek to detect $\gamma$-ray emission from annihilation of dark matter particle candidates \citep[e.g.,][]{Strigari2007,Bonnivard2015}, thus impacting two important fields of astrophysical research.\par

In addition, we show that this approach to studying the tidal disruption of resolved stellar systems by using RR Lyrae stars as tracers to eliminate problems of foreground contamination works for objects with small stellar populations. This approach will not be appropriate for very small UFDs, where fewer than three RR Lyrae stars may be present \citep[e.g.,][for Ursa Major II, Coma Berenices, and Segue 2 and Segue 3, respectively]{Musella2009,DallOra2012,Boettcher2013}, or UFDs closer to the Galactic center where the field density of RR Lyrae stars is significant. But for distant Milky Way satellites, this approach seems promising. 

\acknowledgments
We would like to thank Ian McGreer, Xiaohui Fan, and Linhua Jiang for their assistance with acquiring our Bok observations in coordination with the Beijing-Arizona Sky Survey (BASS). The BASS is a collaborative program between the National Astronomical Observatories of Chinese Academy of Science and Steward Observatory of the University of Arizona. It is a key project of the Telescope Access Program (TAP), which has been funded by the National Astronomical Observatories of China, the Chinese Academy of Sciences (the Strategic Priority Research Program “The Emergence of Cosmological Structures” Grant No. XDB09000000), and the Special Fund for Astronomy from the Ministry of Finance. The BASS is also supported by the External Cooperation Program of Chinese Academy of Sciences (Grant No. 114A11KYSB20160057) and Chinese National Natural Science Foundation (Grant No. 11433005).

CG, BW, and JLC are partially supported by NSF Faculty Early Career Development (CAREER) award AST-1151462. DJS acknowledges support from NSF grant AST-1412504. JS acknowledges support from NSF grant AST-1514763 and a Packard Fellowship.

\facilities{Blanco (DECam), Bok (90Prime)}
\software{Astropy \citep{astropy}, \href{http://dx.doi.org/10.5281/zenodo.14833}{GATSPY} \citep{Vanderplas2015}, Matplotlib \citep{Matplotlib}, NumPy \citep{numpy}, \href{http://www.stsci.edu/institute/software_hardware/pyraf}{PyRAF}, SciPy \citep{scipy}}

\bibliographystyle{aasjournal}
\bibliography{library}

\appendix
\renewcommand\thefigure{\thesection.\arabic{figure}}

\section{Additional Light Curves}
\label{appendix:lightcurves}
\setcounter{figure}{0}

We present light curves for the variables found in \cite{Musella2012} using our CTIO DECam and Bok 90Prime photometry. We generally reproduce the periods of \cite{Musella2012} to within 1\% accuracy, except for V3, where there is a 2\% discrepancy. We note that in our \emph{r} band light curve there are observations at maximum light that appear to be considerably brighter than the best-fit template predicts. However, our light curve for V3 appears to have less scatter overall than theirs, so our period may be more reliable. For details on period determination, see \S \ref{section:selection}.\par
We would also like to mention that our light curve for V8 has poor phase coverage, including virtually no coverage at maximum light, despite having many detections in our photometry. We reproduce the period found by \cite{Musella2012}, but we would caution readers to take our amplitude and average magnitude measurements for V8 with a grain of salt.\par

\begin{figure}[ht]
  \begin{center}
  \begin{tabular}{cc}
	\includegraphics[width=.4\linewidth, page=8]{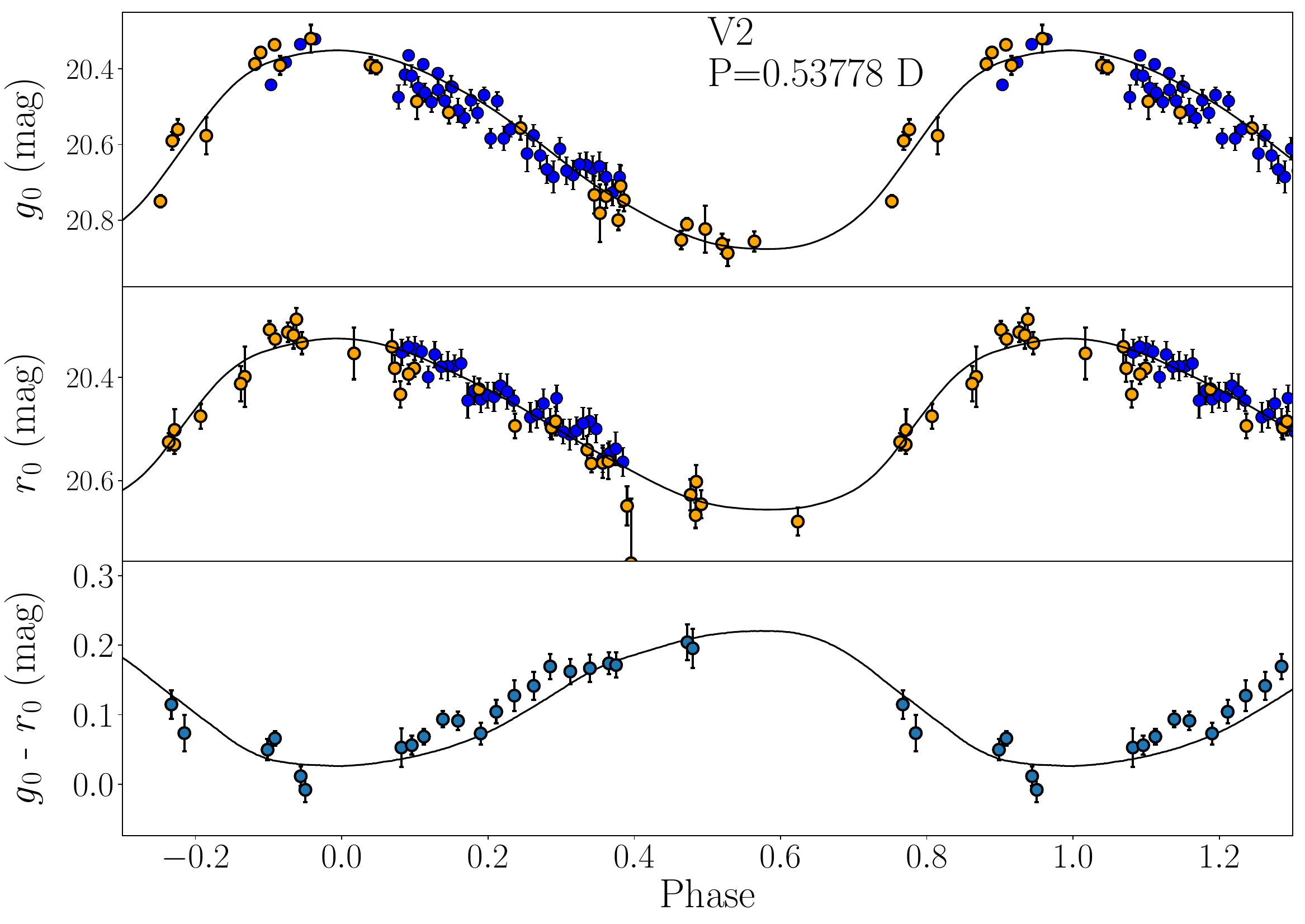} & \includegraphics[width=.4\linewidth, page=14]{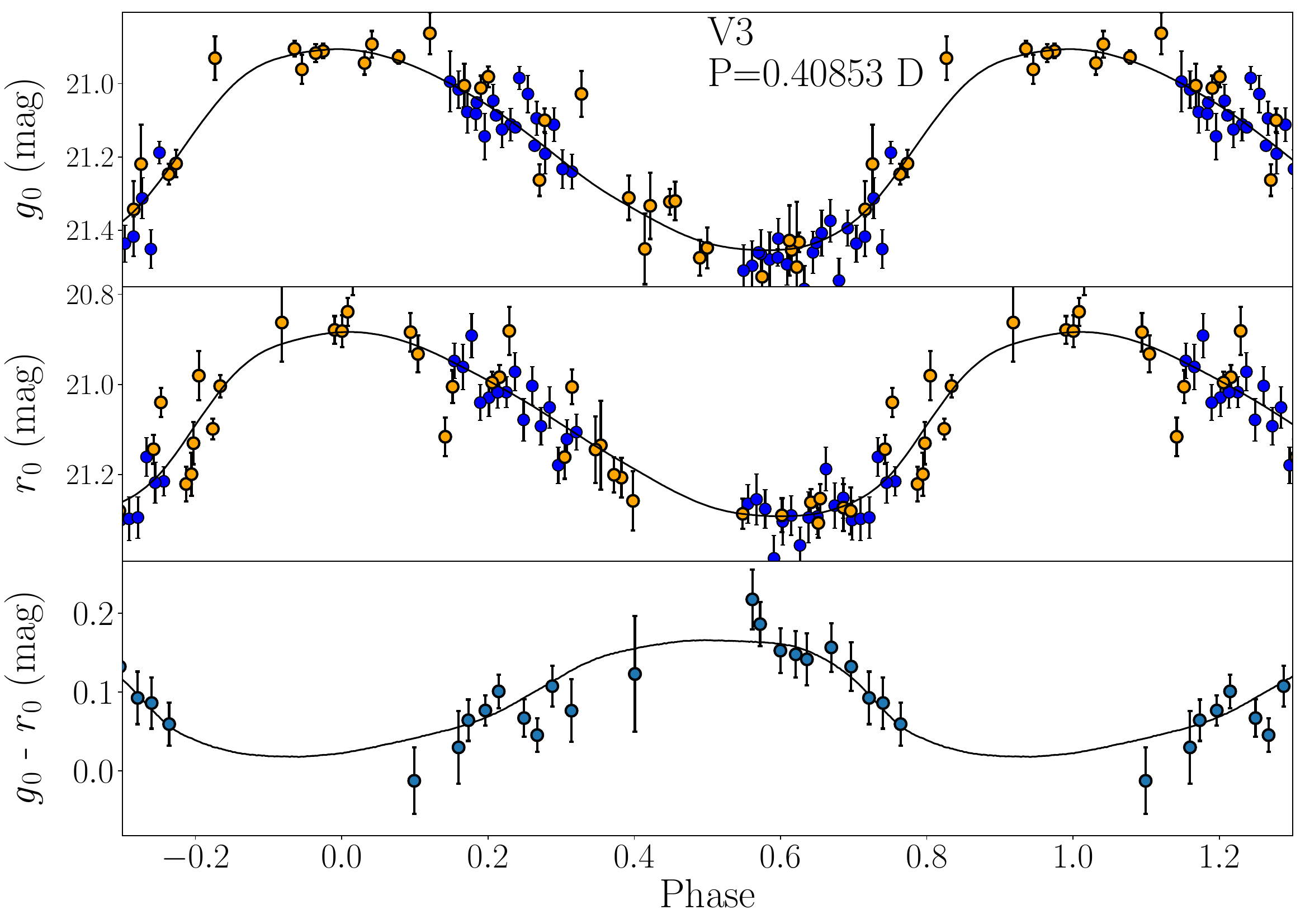} \\
	\includegraphics[width=.4\linewidth, page=32]{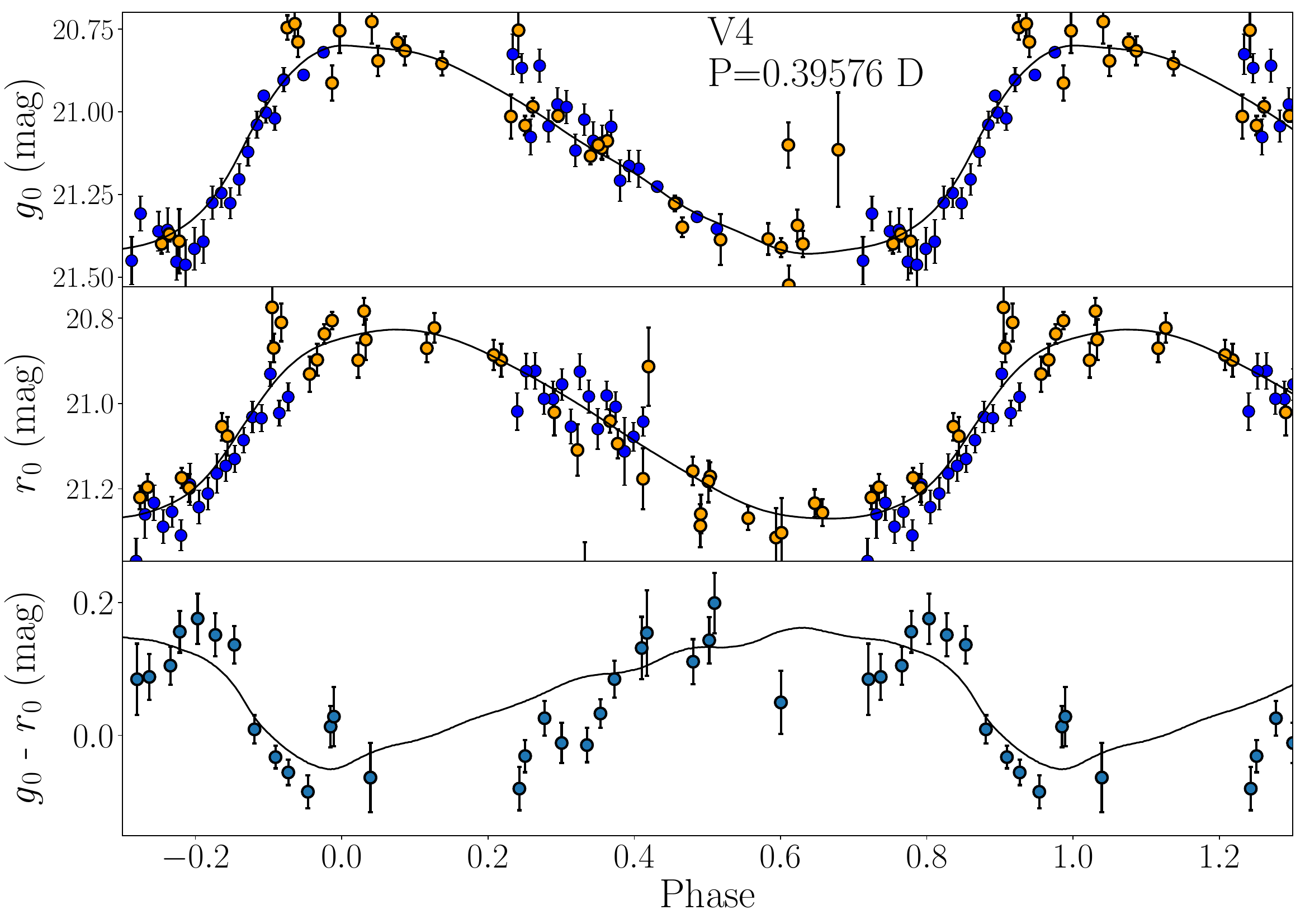} & \includegraphics[width=.4\linewidth, page=38]{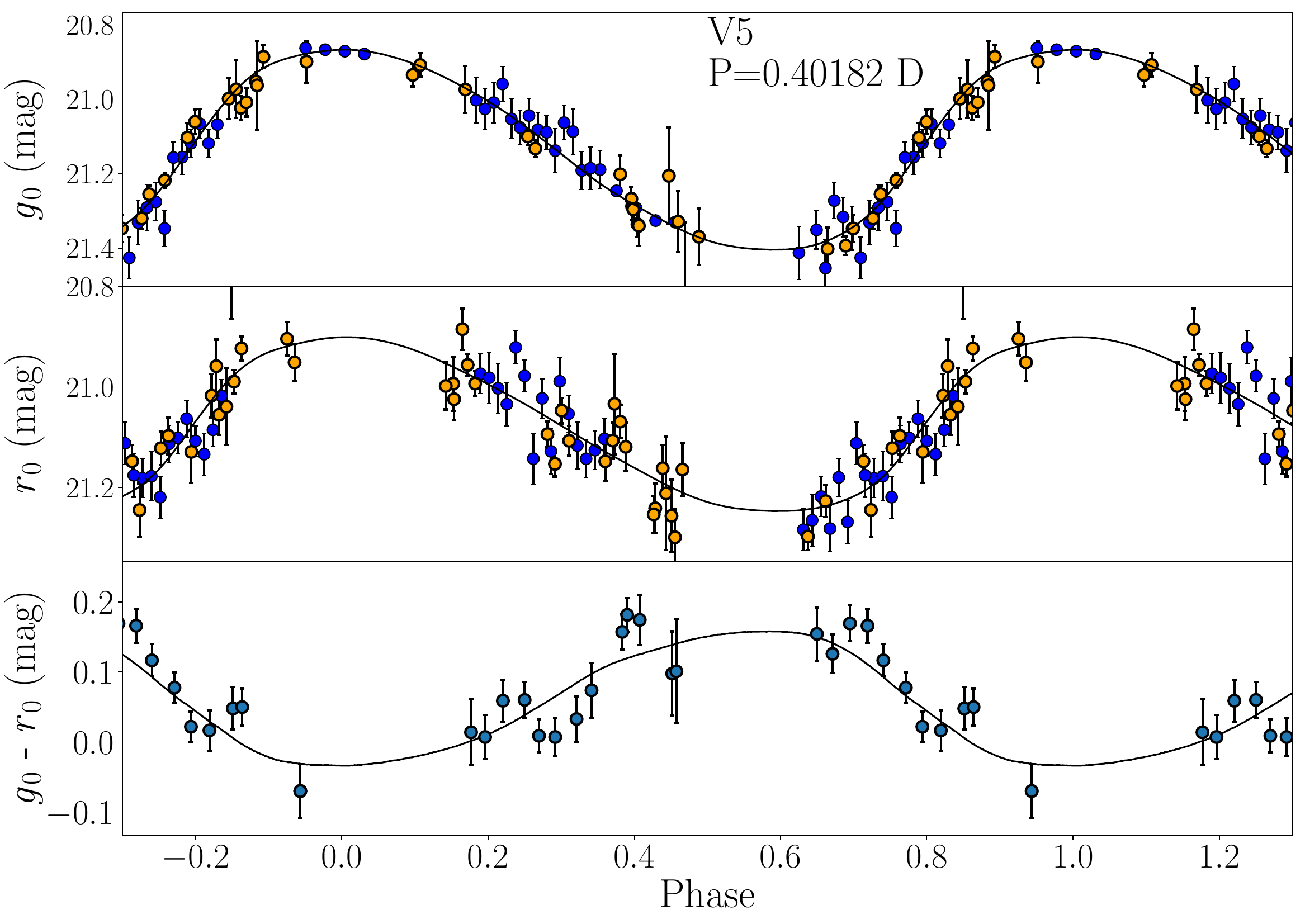} \\
	\includegraphics[width=.4\linewidth, page=44]{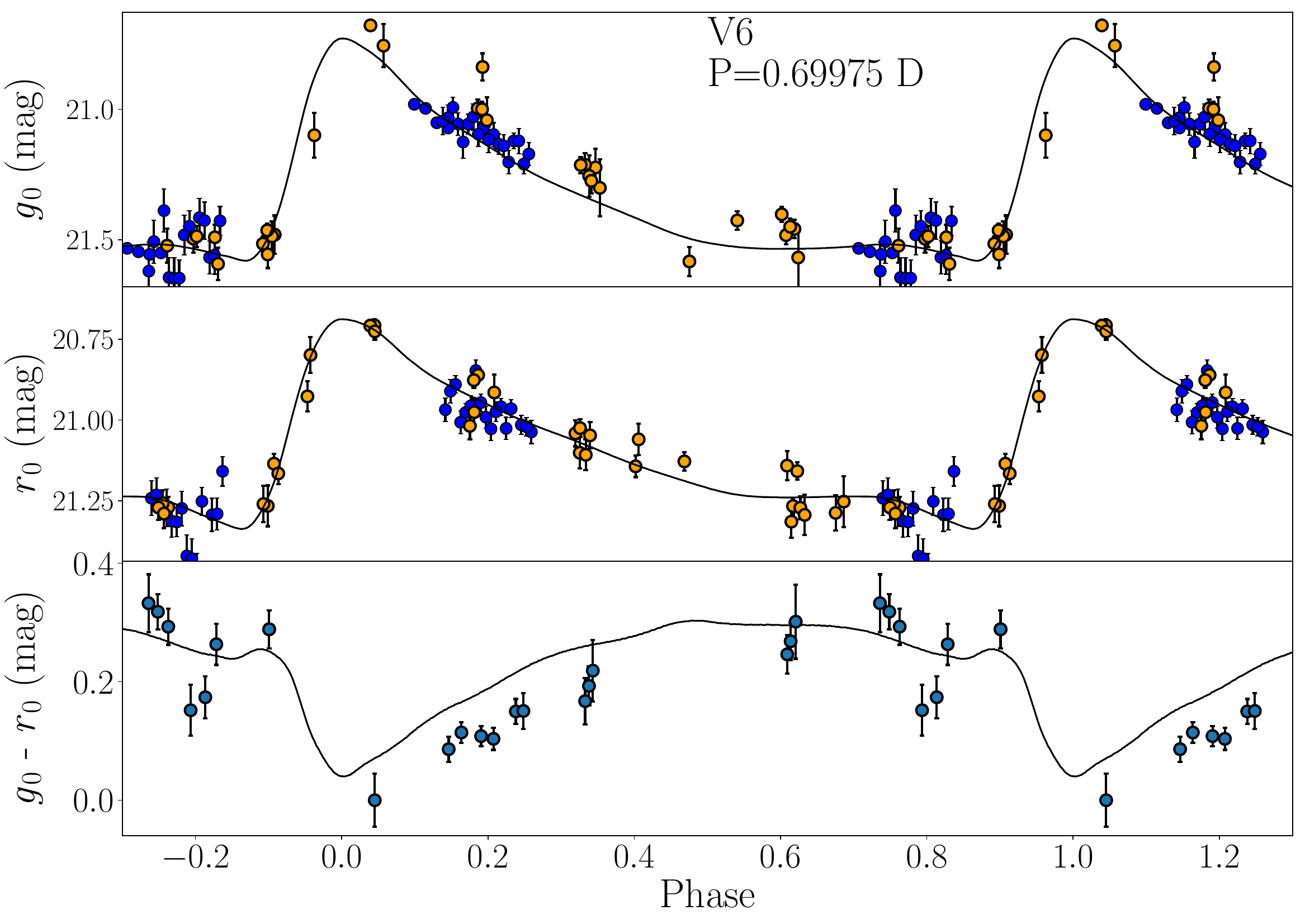} & \includegraphics[width=.4\linewidth, page=28]{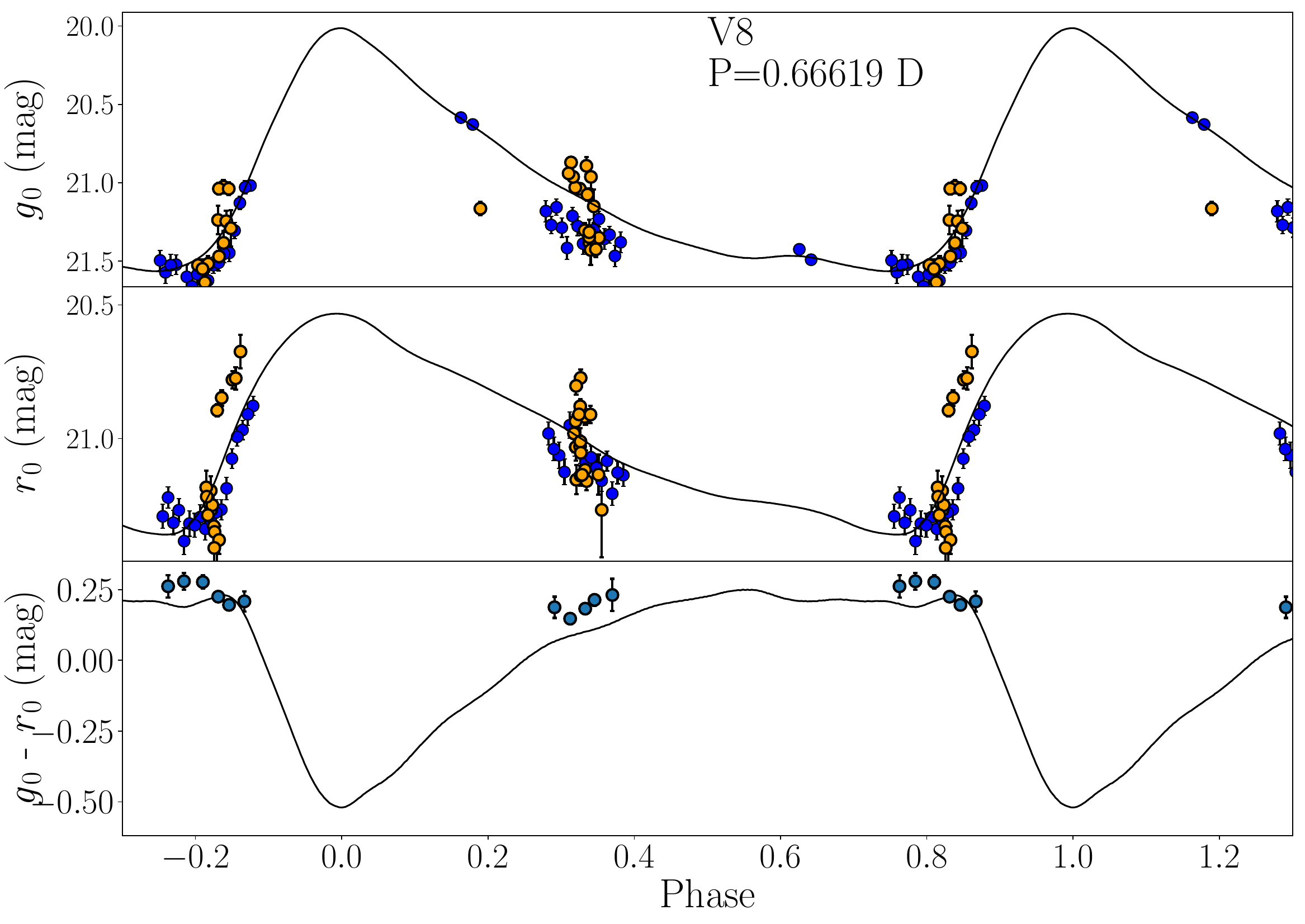} \\
        \includegraphics[width=.4\linewidth, page=20]{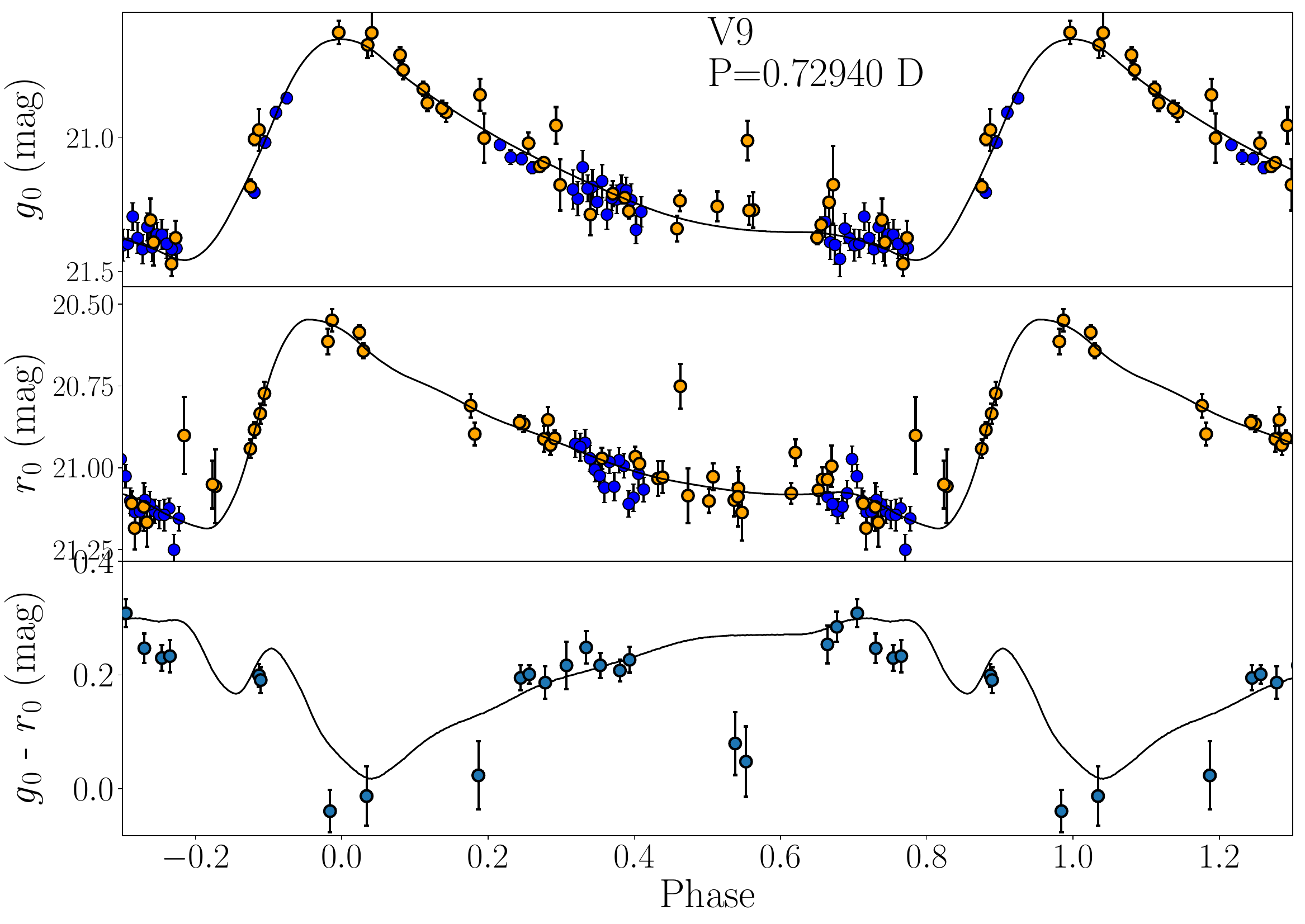} & \includegraphics[width=.4\linewidth, page=2]{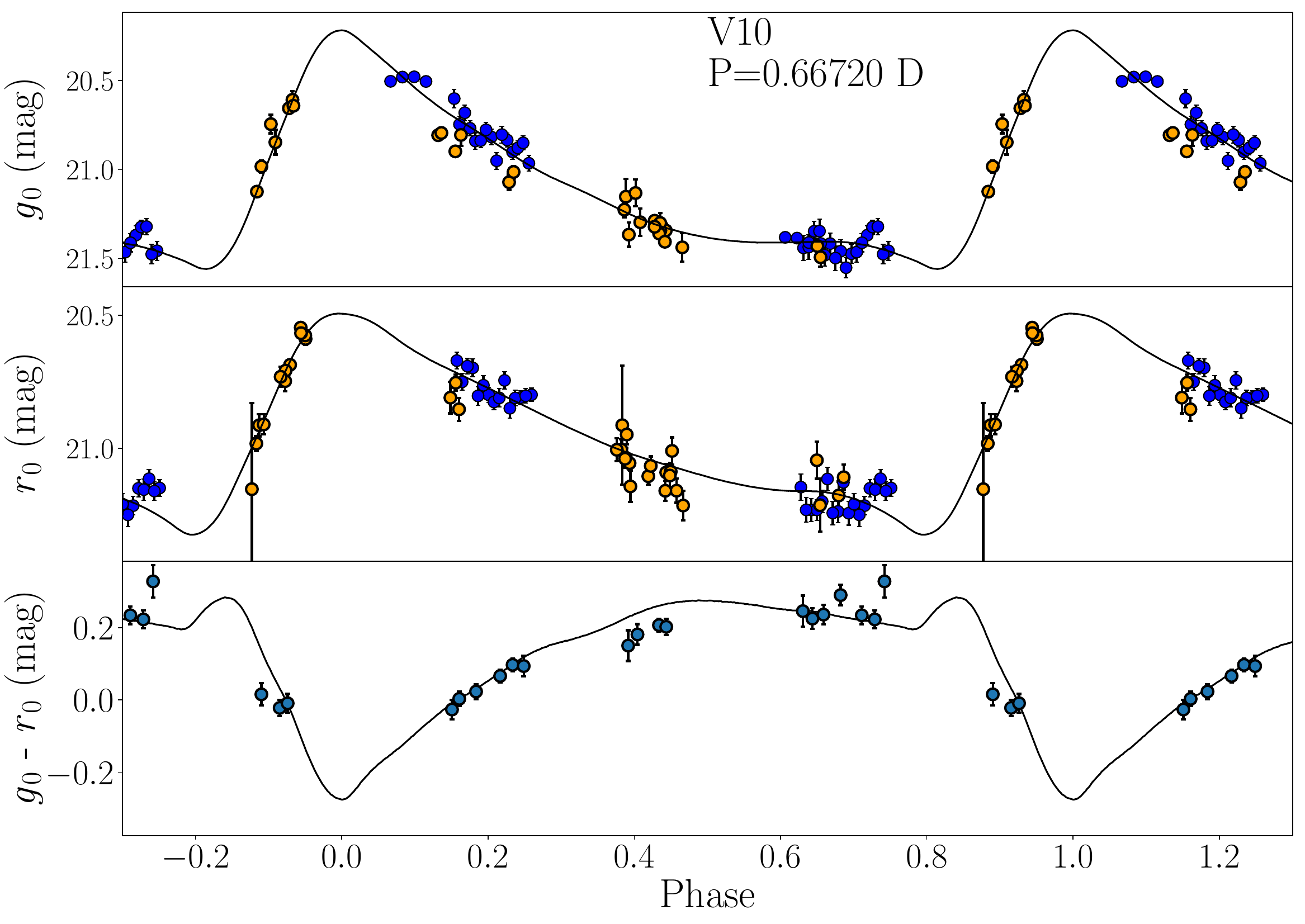}
  \end{tabular}
  \end{center}
  \caption{Period-folded, phase-binned light curves for the RR Lyrae stars from \cite{Musella2012} found in our data. Two of their RR Lyrae stars fell in chip gaps of our DECam discovery data. We used the multi-band periodogram described in \cite{Vanderplas2015} coupled with fits to the templates from \cite{Sesar2010} to characterize the periods, amplitudes, and flux-averaged magnitudes of our variables. DECam observations are in blue and Bok observations are in orange. To generate the \emph{g}\textsubscript{0}-\emph{r}\textsubscript{0} data, we bin our observations such that each point on the light curve is derived from no less than two observations with phase separation no greater than 0.05.}
  \label{extralightcurves1}
\end{figure}

\end{document}